\documentclass[11pt]{article}

\renewcommand{\theequation}{\arabic{section}.\arabic{equation}}
\def\lb{\label}
\def\bb{\bibitem}
\def\be{\begin{equation}}
\def\ee{\end{equation}}
\def\ba{\begin{eqnarray}}
\def\ea{\end{eqnarray}}
\def\dfrac{\displaystyle\frac}
\def\nn{\nonumber}
\def\E{{\cal E}}

\def\C{{\cal C}}
\def\l{\overline{l}}
\def\b{\overline{b}}
\begin{document}

\begin{titlepage}
\date{16 October 2017}
\title{
Motion of charged particles in a NUTty Einstein-Maxwell spacetime\\ and
causality violation}

\author{G\'erard Cl\'ement$^a$\thanks{Email: gclement@lapth.cnrs.fr},
Mourad Guenouche$^{a,b,c}$\thanks{Email: mourad.guenouche@lapth.cnrs.fr} \\ \\
$^a$ {\small LAPTh, Universit\'e Savoie Mont Blanc, CNRS, 9 chemin de Bellevue,} \\
{\small BP 110, F-74941 Annecy-le-Vieux cedex, France} \\
$^b$ {\small Laboratoire de Physique Th\'eorique, D\'epartement de Physique,} \\
{\small Facult\'e des Sciences Exactes, Universit\'e de Constantine 1, Algeria} \\
$^c$ {\small Department of Physics, Faculty of Sciences, Hassiba
Benbouali University of Chlef, Algeria}}

\maketitle

\begin{abstract}
We investigate the motion of electrically charged test
particles in spacetimes with closed timelike curves, a subset of
the black hole or wormhole
Reissner-Nordstr\"om-NUT spacetimes without periodic identification of
time. We show that, while in the wormhole case there are closed
worldlines inside a potential well, the wordlines of
initially distant charged observers moving under the action of the Lorentz force
can never close or self-intersect.
This means that for these observers causality is preserved, which is an instance
of our weak chronology protection criterion.
\end{abstract}
\end{titlepage}
\setcounter{page}{2}

\setcounter{equation}{0}
\section{Introduction}
A variety of solutions to the equations of general relativity
describe spacetimes with closed timelike curves (CTCs). The common
view regarding these is that they violate causality, ``for, one
could imagine that with a suitable rocketship one could travel round
such a curve'' \cite{hawell} and eventually return to one's original
spacetime position after a finite proper time lapse, thus opening
the possibility for time travel. In \cite{haw92}, Hawking examined
this possibility and gave a number of arguments which strongly
support the chronology protection conjecture: ``The laws of physics
do not allow the appearance of closed timelike curves''.

Basically, what Hawking proved is that the creation of CTCs in a
finite region of an initially causal (CTC-less) spacetime is
classically forbidden by the average weak energy condition. If this
is waived, then quantum back-reaction effects come to the rescue to
prevent the appearance of CTCs. However these arguments do not rule
out the possibility of spacetimes with eternal CTCs. Such a
possibility was first considered by G\"odel \cite{godel}. CTCs exist
in the inner region ($r<0$) of the Kerr or Kerr-Newman spacetimes.
Other stationary solutions of the Einstein-Maxwell equations which
present CTCs are the Taub-NUT spacetime \cite{taub,NUT} and its
electromagnetic extensions, the Reissner-Nordstr\"om-NUT or Brill
spacetimes \cite{brill}. These are characterized by a gravimagnetic
or NUT charge and, similarly to the case of the magnetic monopole
with its Dirac string, present one or two line metric singularities
known as Misner strings. These can be removed if time is
periodically identified with a period proportional to the NUT charge
\cite{misner}. However there are then CTCs everywhere in the
stationary sector, so that the spacetime is definitely acausal.
Another drawback of this periodicity in time is that the spacetime
cannot be consistently extended beyond the second (interior)
horizon. The other option is to abstain from this periodic
identification, and so to retain the Misner string singularities. As
implied in \cite{ZS}, and shown explicitly in \cite{TNUT} and
\cite{wnut}, the Brill spacetimes are then geodesically complete,
i.e. the Misner string singularities do not show up in the geodesic
motion. However there are still CTCs in a stationary neighborhood of
the Misner strings.

The point of view advocated in \cite{TNUT,wnut} is that the presence
of closed timelike curves does not in itself give rise to causality
violations, which can arise only if it is possible for an observer
to follow such CTCs. Freely falling observers follow geodesics, so a
necessary condition for causality preservation should be that the
spacetime should be free from closed timelike geodesics (CTGs).
These have been shown to be absent from the Taub-NUT spacetime
(without time periodicity) in \cite{TNUT}, and from the Brill
spacetimes in \cite{wnut}, provided the parameter fixing the
strength of the Misner strings is chosen in a certain range. Another
class of observers are not freely falling because they carry
electric charge, and so travel in an Einstein-Maxwell spacetime as
charged test particles following the Lorentz force law. Causality
will be preserved if their worldlines are not closed or
self-intersecting. Such a question, i.e. the possibility for a
charged particle to move along a CTC in the G\"odel universe under
the effect of a weak sourceless magnetic field, was previously
addressed in \cite{novello}. The existence of closed wordlines
(CWLs) of a charged particle was investigated in \cite{wnut} in the
case of a very special Brill spacetime. It was found that CWLs do
occur inside a potential well, but it was argued that the worldlines
followed by charged observers which are initially distant (and thus
outside the potential well) will necessarily be causal. Finally,
more adventurous observers could try to follow CTCs by using a
``suitable rocketship'', but one can argue that the (classical)
back-reaction of the energy expended to follow a non-geodesic path
(in the case of an uncharged spaceship) would ultimately be so large
that it would deform the background spacetime geometry in such a way
as to preserve chronology. If the arguments concerning these various
instances of observer wordlines prove to be correct, then we can
formulate the weak chronology protection criterion (WCP):

{\em Spacetimes with closed timelike curves do not classically
violate causality, if no worldline followed by an initially distant
observer can possibly be closed or self-intersecting.}

The purpose of the present paper is to study the motion of charged
particles, with the question of CWLs in mind, in a class of Brill
spacetimes more general than that considered in \cite{wnut}. This
motion was recently analyzed in \cite{cos} in the case of the
electric Kerr-Newman-NUT spacetime, but the question of causality
was not investigated there. Here, in order to have a simple
tractable form for the effective potential, we restrict to the case
of the massless magnetic Reissner-Nordstr\"om-NUT spacetime,
considering the whole range of black hole, extreme black hole and
wormhole solutions.

In the next section, we summarize the general results of \cite{wnut}
concerning the motion of an electrically charged test particle in a
Brill spacetime. We then specialize to the case of the massless and
electrically neutral spacetimes (with only magnetic and NUT charges)
in Sect. 3, where we analyze the possible circular orbits. This
analysis is used in Sect. 4 to prove that, for this class of Brill
spacetimes, the Lorentz-force motion of an initially distant charged
observer is always causal. Our conclusions are summarized in the
last section.

\setcounter{equation}{0}
\section{Charged particle motion in the Brill spacetime}

The Reissner-Nordstr\"om-NUT solution of the Einstein-Maxwell system of
equations is given by
 \ba\lb{nworm}
ds^2 &=& - f(dt-2n(\cos\theta+C)\,d\varphi)^2 + f^{-1}dr^2 +
(r^2+n^2)(d\theta^2 + \sin^2\theta d\varphi^2)\,, \nn\\
A &=& \Phi(dt-2n(\cos\theta+C)\,d\varphi)\,,
 \ea
with
 \ba\lb{fem}
f &=& \frac{(r-m)^2+b^2}{r^2+n^2}\,, \quad \Phi = \frac{qr +
p(r^2-n^2)/2n}{r^2+n^2}\,, \\ && (b^2 = q^2+p^2-m^2-n^2)\,. \nn
 \ea
This solution depends on four parameters associated with conserved
charges, the mass $m$, the NUT or gravimagnetic charge $n$, the
electric charge $q$ and the magnetic charge $p$. We assume here
$n\neq0$. In this case, if time is not periodically identified the
maximal analytic extension of the spacetime (\ref{nworm}) is geodesically complete
\cite{wnut}, and corresponds to a black hole for $b^2<0$, to an
extreme black hole for $b^2=0$ and to a traversable Lorentzian
wormhole for $b^2>0$. 

The above solution was first given (in the Taub form) for $C=0$
by Brill \cite{brill}. An unphysical local coordinate transformation 
$t\to t-2nC\varphi$ generates the family of solutions (\ref{nworm})
depending on the additional parameter $C$, which we shall refer to
as Brill spacetimes for simplicity. As shown in
\cite{wnut}, for any $|C|>1$ there are Brill spacetimes with closed
null geodesics, so to ensure the possibility of weak chronology
protection we restrict to $|C|\le1$. The parameter $C$ governs the
strength of the Misner string metric singularities at $\theta=0$ 
and/or $\pi$. The commonly preferred values
are $C=1$ (singularity at $\theta=0$), $C=-1$ (singularity at
$\theta=\pi$), or $C=0$ (two symmetrical singularities at $\theta=0$
and $\theta=\pi$).

The equations of motion of of a charged particle with mass $m_c$ and
charge $q_c$ may be derived from the Lagrangian
 \be\lb{lag}
L = \frac12\,g_{\mu\nu}\dot{x}^\mu\dot{x}^\nu + \kappa
A_\mu\dot{x}^\mu,
 \ee
where $\dot{}= d/d\tau$, with $\tau$ the proper time, and
$\kappa=q_c/m_c$. The momenta $p_\mu = g_{\mu\nu}\dot{x}^\nu +
\kappa A_\mu$ conjugate to the cyclic variables $t$ and $\varphi$
are constants of the motion, $p_t=-E$ and $p_\varphi=J_z+2nCE$.
Noting that the electromagnetic gauge chosen in (\ref{nworm}) is
such that $A^\varphi=0$, we obtain the equations of motion for the
time and azimutal coordinates:
 \ba
&& \dot{t} - 2n(\cos\theta+C)\,\dot\varphi = f^{-1}\E    \qquad
\left[\E(r) = E + \kappa\Phi(r)\right], \lb{t}\\
&& \dot\varphi = \frac{J_z -
2nE\cos\theta}{(r^2+n^2)\sin^2\theta}\,. \lb{varphi}
 \ea

This last equation is the same as in the case of a neutral particle
\cite{ZS}. Also, the part of the Lagrangian (\ref{lag}) which
depends explicitly on the coordinate $\theta$ is
 \be
L_\theta = \frac12g_{\varphi\varphi}\dot\varphi^2 +
g_{t\varphi}\dot\varphi(\dot{t}+\kappa A^t),
 \ee
with $A^t = -\Phi/f$. Owing to (\ref{t}), this actually does not
depend on $\kappa$ \cite{wnut}, so that the equation of motion for
$\theta$,
 \be
  [(r^2+n^2)\dot\theta]\,\dot{} = (r^2+n^2)\sin\theta\cos\theta
\,\dot\varphi^2 - 2nE\sin\theta\,\dot\varphi\,,
 \ee
is also the same as for a neutral particle. The angular equations of
motion can be first integrated \cite{ZS} to
 \be\lb{J}
\vec{L} + \vec{S} = \vec{J},
 \ee
meaning that the constant total vector angular momentum
$\vec{J}=(J_x,J_y,J_z)$ is the sum of the usual orbital angular 
momentum (rescaled by a factor $(r^2+n^2)/r^2$)) $\vec{L} = 
(r^2+n^2)\,\hat{r}\wedge\dot{\hat{r}}$, where $\hat{r}$ is a unit
vector normal to the two-sphere, and a NUT charge contribution 
$\vec{S}=2nE\hat{r}$. It follows from the orthogonality
of $\vec{L}$ and $\vec{S}$ that
 \be\lb{para}
\vec{J}\cdot\hat{r}=2nE\,.
 \ee
so that \cite{ZS} the spherical sections $r =$ constant of the
orbits are small circles (parallels) $\C$ with polar axis $\vec{J}$
and colatitude $\eta=\arccos(2nE/J)$, where $J^2 \equiv \vec{J}^2$.
Squaring (\ref{J}) leads to
 \be\lb{J2LE}
\vec{L}^2 = l^2 \equiv J^2 - 4n^2E^2\,,
 \ee
which can be rewritten as
 \be\lb{l2}
(r^2+n^2)^2[\dot\theta^2 + \sin^2\theta\,\dot\varphi^2] = l^2\,.
 \ee
As discussed in \cite{kagra}, the system of equations (\ref{varphi})
and (\ref{l2}) can be completely integrated by introducing the Mino
time $\lambda$, related to the proper time by
\begin{equation}
d\tau =\left( r^{2}+n^{2}\right) d\lambda\,.   \label{mint}
\end{equation}
We only give here the solution for $\theta(\lambda)$:
 \be\lb{theta2}
\cos\theta = \cos\psi\cos\eta + \sin\psi\sin\eta\cos(J\lambda)\,,
 \ee
with $\psi=\arccos(J_z/J)$. This shows that the Mino period of the
angular motion is $2\pi/J$.

The time evolution can be split \cite{kagra} according to
 \be
t(\lambda) = t_r(\lambda) + t_{\theta}(\lambda)\,,
 \ee
where the radial and angular contributions to $t(\lambda)$ solve the
equations
 \ba
\frac{dt_r}{d\lambda} &=& (r^2+n^2)\frac{\E(r)}{f(r)}\,, \lb{tr}\\
\frac{dt_{\theta}}{d\lambda} &=& \frac{2n(\cos\theta+C)(J_z -
2nE\cos\theta)}{\sin^2\theta}\lb{ta}\,.
 \ea
The solution to equation (\ref{tr}) depends on the solution of the
equation for radial motion (\ref{rad}). The solution to equation
(\ref{ta}) is, up to an additive constant,
 \ba\lb{solta}
t_{\theta}(\lambda) &=& 4n^2E\lambda + 2n(C+1)
\arctan\left[\frac{\cos\psi-\cos\eta}{1-\cos(\psi-\eta)}
\tan\,\frac{J\lambda}2\right] \nn\\
&+& 2n(C-1)\arctan\left[\frac{\cos\psi+\cos\eta}{1+\cos(\psi-\eta)}
\tan\,\frac{J\lambda}2\right]\,,
 \ea
in the interval $-\pi/J < \lambda < \pi/J$.

Finally, after eliminating in the Hamiltonian $H$ associated with
(\ref{lag}) the angular and temporal velocities in terms of the
constants of the motion $l$ and $E$, and conventionally normalizing
proper time by setting $H=-1/2$, we obtain the effective radial
equation \cite{wnut}
 \be\lb{rad}
\dot{r}^2 + W(r)=0,\,\quad \left[W(r) \equiv
f(r)\left(1+\frac{l^2}{r^2+n^2}\right) - \E^2(r)\right]\,,
 \ee
where $\E(r)$ has been defined in (\ref{t}). In the stationary
sector ($f(r)>0$), $\E^2(r)$ must remain positive, so that the sign
of $\E(r)$ is a constant of the motion. From (\ref{tr}), a necessary
condition for coordinate time and proper time to have the same
orientation is $\E(r)>0$. This condition, which ensures that the
radial contribution $t_r(\lambda)$ is an increasing function, is not
sufficient, because the angular contribution $t_\theta(\lambda)$ is
not generically increasing. Indeed, as shown in \cite{wnut}, for
parallels $\C$ which do not circle a Misner string,
$t_\theta(\lambda)$ is in the large a decreasing function of Mino
time for all values of the parameter $C$. So it could be possible
for the time coordinate $t=t_r+t_\theta$ to take again the same
value after a finite lapse of Mino or proper time. If after this
lapse the radial and angular coordinates also took again the same
values, then the wordline would be closed (or self-intersecting). In
order to investigate this possibility, it is first necessary to
discuss the possible radial motions determined by (\ref{rad}).

\setcounter{equation}{0}
\section{Circular orbits: case of the massless magnetic Brill spacetime}

The discussion of the radial motion depends on the position of the
stationary points of the effective potential $W(r)$,
 \be\lb{circ}
W(r) = W'(r) = 0.
 \ee
which correspond to circular orbits. The equation $W(r)=0$ is
generically of fourth order \cite{cos}. However it can be reduced to
a second-order equation if $W(r)$ is even in $r$, i.e. in the
special case $m=0$, $q=0$ of a
massless magnetic Brill spacetime, to which we now restrict. The
effective energy can then be written
 \be\lb{effen}
{\cal E}(r) =  E_0 + \frac\beta2 y \qquad
\left(0<y=\frac{n^2}{r^2+n^2}\le1\right),
 \ee
with $\beta=-2\kappa p/n$, and
 \be\lb{E0E}
E_0 = \E(\infty) = E-\beta/4.
 \ee
Note that the choice $\E(r)>0$ implies $E_0>0$ in the case of orbits
extending to infinity, i.e. if $E_0^2>1$. On account of
(\ref{effen}) the effective potential (\ref{rad}) reduces to
 \be
W(y) = Ay^2+By+D
 \ee
with
 \be
A=\alpha\l^2-\beta^2/4,\;\; B=\alpha+\l^2-\beta E_0,\;\;D=1-E_0^2,
 \ee
where we have put $\l=l/n$, $\alpha=\b^2-1$ with $\b^2=b^2/n^2 \ge
-1$ (the very special case $p^2=n^2$ treated in \cite{wnut}
corresponds to $\alpha=0$).

\subsection{Circular orbits with $r=0$}

The equation $W'(r) = 0$ has two solutions. The first is $r=0$, with
effective energy ${\cal E}(0) = E_0 + \beta/2$ given by the positive
root of
 \be
W(y=1) = A+B+D = (\alpha+1)(\l^2+1) - {\cal E}(0)^2 = 0.
 \ee
The corresponding orbit is stable or unstable, depending on whether
the second derivative
 \be W''_r(0) = -\frac2{n^2}\,W'_y(1) = -\frac2{n^2}\,(2A+B) =
-\frac2{n^2}\,\left[\alpha + \l^2 + 2\alpha\l^2 -
\beta{\E}(0)\right]
 \ee
is positive or negative. In the black-hole or extreme-black-hole case, 
$-2\le\alpha\le-1$, $W(0)<0$ for all $\E(0)>0$, so that $r=0$ cannot 
be a stationary point. In the wormhole case $\alpha>-1$, the effective energy is
 \be\lb{Ebga}
{\cal E}(0) = \b\gamma \qquad \left(\b \equiv \sqrt{\alpha+1},\;\;
\gamma \equiv \sqrt{\l^2+1}\right).
 \ee
If
 \be\lb{stab0}
\beta > \frac{\alpha + \l^2 + 2\alpha\l^2}{\b\gamma} =
\alpha\frac\gamma\b + \l^2\frac\b\gamma = 2\b\gamma - \frac\gamma\b
- \frac\b\gamma,
 \ee
$r=0$ is a minimum of $W$ (stable orbit). On the contrary, if the
inequality (\ref{stab0}) is reversed, $r=0$ is a maximum and this
orbit is unstable.

\subsection{Circular orbits with $r\neq0$}
The second solution is $dW/dy=0$, i.e. $y=y_0$, with
 \be\lb{y0}
y_0=-\frac{B}{2A}.
 \ee
Then $W=0$ if $B^2=4AD$, i.e.
 \be\lb{circn0}
4\alpha\l^2 E_0^2 - 2\beta(\alpha+\l^2)E_0 + \beta^2 +
(\alpha-\l^2)^2 = 0,
 \ee
with discriminant
 \be
\Delta = (\beta^2-4\alpha\l^2)(\alpha-\l^2)^2.
 \ee
This is clearly non-negative if $\alpha\le0$. If $\alpha>0$,
the discriminant is non-negative provided
 \be\lb{Aneg}
\beta^2 \ge 4\alpha\l^2
 \ee
($A\le0$). Then (\ref{circn0}) is solved by
\begin{equation}\lb{E0}
E_{0}=\frac{\beta (\alpha+\l^{2}) - \delta(\alpha-\l^{2})}
{4\alpha\l^{2}},
\end{equation}
where we have put
 \be\lb{da}
\delta \equiv {\rm sign}(\alpha-\l^2)\sqrt{\beta^2-4\alpha\l^2}.
 \ee
Further defining
 \be\lb{ea}
z \equiv \frac{\beta-\delta}{2\l^2},
 \ee
we can rewrite (\ref{E0}) as
 \be\lb{E01}
E_0 = \frac{1+z^2}{2z}.
 \ee
It follows from (\ref{E01}) that $E_0^2>1$ (the limiting value
$E_0^2=1$ for $z^2=1$ generically corresponding from (\ref{y0z}),
see the Appendix, to an orbit at infinity), so that $W(r_0) = 0 >
W(\infty) = 1 - E_0^2$, i.e. the extremum at $r=r_0$, if it exists,
is a local maximum (another way to see this is to compute $W''_r(\pm
r_0) = (4r_0^2y_0^4/n^4)W''_y(y_0) = (8r_0^2y_0^4/n^4)A \le 0$).

From its definition, $y_0=n^2/(r_0^2+n^2)$ given by (\ref{y0})
must be positive and smaller than 1. As shown in the Appendix,
this means that circular orbits with $r_0\neq0$ exist only if the
test particle charge-to-mass ratio $\kappa$ and scaled orbital
angular momentum $\l$ satisfy the relations
 \be
\frac{\alpha + \l^2 + 2\alpha\l^2}{\b\sqrt{\l^2+1}} < -\frac{2\kappa p}{n} < \alpha +
\l^2
 \ee
(the lower bound being irrelevant in the black-hole case $\b^2<0$).

In the very special limiting case $\l^2=\alpha = \beta/2$,
(\ref{da}) and (\ref{ea}) lead to $z=1$, so that also $E_0=1$, and $A=B=D=0$,
leading to $W(r) \equiv 0$. This means that a charged test particle
with parameters fine tuned to those of the Brill spacetime so that
$\kappa \equiv q_e/m_e = (2n^2-p^2)/np$ can follow a (metastable)
circular orbit with any given radius.

\setcounter{equation}{0}
\section{Causality}
Knowing the stationary points of the effective potential, one can
classify the possible worldlines of a charged test particle. We
shall not go here into the details of this analysis, which leads to
results qualitatively similar to those discussed in \cite{wnut} for
$\alpha=0$. The possible orbits are contained in the range of values
of $r$ such that $W(r)\le0$. This range can be finite (in the
wormhole case), corresponding to bound orbits in the potential well
around the minimum $r=0$ of $W(r)$ if the inequality (\ref{stab0})
is satisfied. Or it can be infinite, with either scattering orbits
reflected on the potential barrier (if the maximum of the effective
potential $W(0)$ or $W(\pm r_0)$ is positive), or traversing orbits
going from $r=+\infty$ to $r=-\infty$ (if the maximum is negative).

The only possibly causality violating wordlines, in the sense of our WCP
criterion, correspond to scattering orbits followed by a charged observer.
These wordlines can self-intersect at an event $M_1$ if:

1) the Mino time delay
 \be
\Delta\lambda =
2\int_{r_{turn}}^{r_1}\left[-W(r)\right]^{-1/2}\frac{dr}{r^2+n^2}
 \ee
(where $r_{turn}$ is the turning point, $W(r_{turn})=0$) is an integer
multiple of the period $2\pi/J$, so that the angular coordinates take again
the same values, and

2) the coordinate time delay
 \be\lb{delay}\Delta t = 2\int_{r_{turn}}^{r_1}\,\frac{dt}{d\lambda}\,
\left[-W(r)\right]^{-1/2}\frac{dr}{r^2+n^2}
 \ee
vanishes. As discussed in \cite{wnut}, for an initially distant
observer this will generically not occur, because at distances
(to the black hole horizon or to the wormhole neck, according to
the sign of $b^2$) large before the characteristic length $n$,
the potentially negative angular contribution $dt_\theta/d\lambda$
(which does not depend on the distance $r$) will easily be
balanced by the positive definite radial contribution $dt_r/d\lambda
\simeq E_0r^2$ (see (\ref{tr}) and (\ref{ta})).

But this argument breaks down if the turning point is close to the
maximum, i.e. if $W'(r_{turn})$ is small. Then the observer can
spend a long proper time near the turning point, making a large
number of turns $N$ before returning to infinity, so that the
integral in (\ref{delay}) will be of the order of
 \be
\Delta t \simeq N\Delta_1 t(r_{max}),
 \ee
where $\Delta_1 t(r_{max}) = \Delta_1 t_r(r_{max}) + \Delta_1 t_\theta$,
the time lapse during a Mino period
$2\pi/J$ for a particle on an unstable circular orbit, is not
obviously positive definite. From (\ref{tr}) and (\ref{solta}), the radial
and angular components of $\Delta_1 t(r)$ are given by
 \ba
& \Delta_1 t(r) &= \frac{2\pi}J(r^2+n^2)\frac{\E(r)}{f(r)}\,,\\
& \Delta_1 t_{\theta} &= 2\pi n\left[\frac{4nE}J  + (C+1){\rm
sgn}(J_z-2nE) + (C-1){\rm sgn}(J_z+2nE) \right]\,,
 \ea
For $|C|\le1$, a lower bound for $\Delta_1 t_{\theta}$ is
 \be
\Delta_1 t_{\theta} \ge 2\pi n\left[\frac{4nE}J - 2\right]\,,
 \ee
the lower bound being attained for all orbits if $C=\pm1$, and for
orbits with $-2nE<J_z<2nE$ for other values of $C$. Thus the lower
bound for the net time lapse during one period is
 \be\lb{lowerbound}
\Delta_1 t(r) \ge \frac{2\pi n^2}{J}\left(\Psi(r)
+4E-2\dfrac{J}n\right) ,
\end{equation}
where
\begin{equation}
\Psi(r) = \frac{\E(r)}{yf(r)} > 0,
\end{equation}
and $E$ is related to $E_0$ by (\ref{E0E}). The lower bound
(\ref{lowerbound}) is not positive definite because $J \ge 2nE$. It
is positive provided\footnote{The positivity condition (\ref{Dpos}),
derived under the assumption $\Psi(r)+4E>0$, is sufficient because
$\Psi(r)+4E\le0$ is possible only if $E<0$, so that
$4(\l^2-E\Psi(r)) > 0 \ge \Psi(r)(\Psi(r)+4E)$.}
\begin{equation}\lb{Dpos}
\Delta(r) = \left(\Psi(r)+4E\right)^2 - \left(\dfrac{2J}n\right)^2 =
\Psi(r)^2 +8E\Psi(r) -4\l^2 > 0
\end{equation}
(where we have used (\ref{J2LE})). We now investigate whether
$\Delta(r)$ can vanish in the two cases of circular orbits ($r=0$
and $r=r_0\neq0$) discussed in Sect. 3.

\subsection{$r=0$}
In this case, $\Psi(0)={\cal E}(0)/\b^2 = \gamma/\b$, $E =
\b\gamma-\beta/4$, leading to
 \be
\Delta(0) = \b^{-2}\left[(1+4\b^2)\gamma^2 - 2\beta\b\gamma +
4\b^2\right].
 \ee
$\Delta(0)$ is positive definite if $\beta<0$. If $\beta>0$,
$\Delta(0)$ vanishes for
 \be\lb{gpm}
\gamma = \gamma_{\pm} \equiv \frac{\b}{4\b^2+1}\left[-\beta \pm
\sqrt{\beta^2 - 4(4\b^2+1)}\right].
 \ee
From (\ref{Ebga}), $\gamma=\gamma_+$ is possible only if
$\gamma_+\ge1$, which is ensured if
 \be
\beta \ge 4\b + \frac1{2\b} \ge 2\sqrt2
 \ee
(with equality for $\b^2=1/8$). For $\beta \ge 4\b + 1/\b$,
$\Delta(0)$ can also vanish for $\gamma=\gamma_-$.

Thus, the wordlines of charged particles circling the wormhole neck
$r=0$ can be closed if their orbital angular momentum matches the
value(s) (\ref{gpm}). However these circular orbits are stable,
because $\Delta(0)=0$ can be written
 \be\lb{D00}
\beta=2\b\gamma + \frac\gamma{2\b} + \frac{2\b}\gamma \ge
2(\b\gamma+1)\,,
 \ee
which is stronger than the stability condition
(\ref{stab0})\footnote{Note that (\ref{D00}) means $\E(\infty) =
\b\gamma - \beta/2 \le -1$, leading to $W(\infty)\le0$, so that the
CWLs at $r=0$ could be accessed by quantum tunnelling from infinity,
but only if negative effective energies were allowed, which we have
excluded.}.

\subsection{$r=r_0$}
Using (\ref{Er0}), we find
\begin{equation}
\Psi(r_0) = \frac{\E(r_0)}{y_0(1+\alpha y_{0})} = \frac1{y_0z}\,.
\end{equation}
Using this together with (\ref{E01}), (\ref{E0E}) and (\ref{bada}),
we obtain\footnote{To transform (\ref{Da1}) into (\ref{Da2}), we
have combined the second and fourth term of (\ref{Da1}) and used
(\ref{y0z}).}
 \ba
\Delta(r_0) &=& \frac1{y_0^2z^2} + \frac{4(1+z^2)}{y_0z^2} +
\frac{2(\alpha+\l^2z^2)}{y_0z^2} - 4\l^2 \lb{Da1}\\
&=& \frac1{y_0^2z^2} + \frac{2(\alpha+\l^2z^2)}{y_0z^2} +
\frac8{y_0} - \frac{4\alpha}{z^2} \lb{Da2} \\
&=& \frac{(1+\alpha y_0)^2}{y_0^2z^2} - \frac{\alpha(\alpha+4)}{z^2}
+ \frac{2(4+\l^2)}{y_0} \,, \lb{Da3}
 \ea
which is positive definite for $-2\le\alpha\le0$.

To cover the complementary range $\alpha\ge0$, we use another
expression of $\Delta(r_0)$, obtained by inserting (\ref{zy0}) into
(\ref{Da1}) or (\ref{Da3}),
 \be
(1+\alpha y_0)y_0^2\Delta(r_0) = 4\alpha \l^2 y_0^2(1-y_0) + 4\alpha
y_0^2 + (2\alpha+3\l^2+8)y_0 + 1,
 \ee
which is positive definite for $y_0 \in [0,1]$ if $\alpha\ge0$.

So the only possible CWLs with circular orbits have $r=0$, but these
have $\E(\infty)<0$ and so cannot attract charged test particles
coming from infinity.

\setcounter{equation}{0}
\section{Conclusion}
We have investigated the motion of electrically charged test
particles in a spacetime with closed timelike curves, the Brill or
Reissner-Nordstr\"om-NUT spacetime with only magnetic and
gravimagnetic (NUT) charges, and without periodic identification of
time. We have argued that causality violations can be observed by an
initially distant charged observer only if he can be attracted by an
unstable closed worldline with circular orbit. It turns out that the
only circular orbits around which the net coordinate time lapse
vanishes, corresponding to a closed wordline, are stable. It follows
that no wordline followed by an initially distant charged observer
moving under the action of the Lorentz force can possibly
self-intersect, meaning that, in this specific dynamical framework,
causality is preserved in the sense of our weak chronology
protection criterion.

This work should be extended in several directions. First, the same
problem should be investigated in the general case of the Brill
spacetime with all four charges non-vanishing. Our guess is that
weak causality should also hold in that case, although a complete
analytical proof seems very difficult, unless methods more powerful
than those of the present paper are used. Also, the possible
existence of closed or self-intersecting wordlines should be
similarly investigated in the case of other spacetimes with CTCs, in
order to see whether they satisfy weak causality. The simplest cases
are presumably those of three-dimensional spacetimes, such as BTZ
\cite{btz}, and warped AdS black hole spacetimes
\cite{tmgbh,adtmg,ALPSS}, which both admit CTCs. The latter are
self-consistent solutions of the three-dimensional Einstein-Maxwell
theory with gravitational and electromagnetic Chern-Simons terms
\cite{ACGL}, so that Lorentz-force motion of charged observers could
be investigated for causality in a fashion similar to that of the
present paper.

\section*{Acknowledgments}
We thank Dmitry Gal'tsov for fruitful discussions and suggestions,
and J\'ulio Fabris for valuable comments. MG acknowledges the
support of the Ministry of Higher Education and Scientific Research
of Algeria (MESRS) under grant 0092009009.

\renewcommand{\theequation}{A.\arabic{equation}}
\setcounter{equation}{0}
\section*{Appendix: Circular orbits with $r\neq0$}
From (\ref{da}) and (\ref{ea}), $\beta$ and $\delta$ can be
expressed in terms of $z$ as
 \be\lb{bada}
\beta = \frac{\alpha + \l^2z^2}z, \quad \delta = \frac{\alpha -
\l^2z^2}z.
 \ee
Using this, we obtain from (\ref{y0})
 \be\lb{y0z}
y_0 = \frac{-\delta(\alpha+\l^{2}) + \beta(\alpha-\l^{2})}
{2\alpha\l^{2}\delta} = \frac{1-z^2}{\l^2z^2-\alpha},
 \ee
 which can be inverted to
 \be\lb{zy0}
z^2 = \frac{1+\alpha y_0}{1+\l^2 y_0}.
 \ee
We also obtain from (\ref{bada}) and (\ref{y0z}) the value of the
effective energy $\E(r_0)=E_0+(\beta/2)y_0$:
 \be\lb{Er0}
\E(r_0) = \frac{(\l^2-\alpha)z}{\l^2z^2-\alpha} = (1+\l^2y_0)z,
 \ee
so that the effective energy is positive provided
 \be
z>0.
 \ee

In the black-hole case or extreme-black-hole case, $-2\le\alpha\le-1$,
(\ref{zy0}) is positive definite provided
 \be\lb{ybh}
0 < y_0 < y_h = - \frac1\alpha,
 \ee
so that the circular orbits must be outside the horizon ($r_0>r_h$).
The allowed range of $z$ is then from (\ref{y0z})
 \be
0 < z < 1,
 \ee
leading from (\ref{bada}) to the condition for the existence of these
circular orbits:
 \be
\l^2 > \beta-\alpha.
 \ee

In the wormhole case, $\alpha>-1$, $y_0$ can vary in the full range
$0<y_0<1$, leading to the allowed range of $z$
 \be\lb{cond}
\begin{array}{lcc}
\dfrac\b\gamma < z < 1 \quad & {\rm if} & \l^2>\alpha, \\
1 < z < \dfrac\b\gamma \quad & {\rm if} & \l^2<\alpha
\end{array}
 \ee
(where $\b$ and $\gamma$ are related to $\alpha$ and $\l^2$ by
(\ref{Ebga})). Both cases lead to the same bounds for the existence
of an unstable circular orbit of radius $r=\pm r_0$,
 \be\lb{synth}
\frac{\alpha + \l^2 + 2\alpha\l^2}{\b\gamma} < \beta < \alpha +
\l^2.
 \ee
For $\alpha>0$, the lower bound ensures that the first existence
condition (\ref{Aneg}) is satisfied, due to the identity
 \be
(\alpha + \l^2 + 2\alpha\l^2)^2 = 4\alpha\l^2\b^2\gamma^2 + (\alpha
- \l^2)^2.
 \ee
Note that in the parameter range (\ref{synth}) there is also from
(\ref{stab0}) a stable circular orbit at $r=0$.

\end{document}